\documentclass[runningheads]{svmult}

\usepackage{makeidx}   
\usepackage{graphicx}  
\usepackage{subeqnar}  
\usepackage{multicol}  
\usepackage{physprbb}  
\makeindex             

\begin{document}
\title*{Radio Properties of EROs \protect\newline in the Phoenix Deep Survey}
\toctitle{Radio Properties of EROs 
\protect\newline in the Phoenix Deep Survey}
%
%
\titlerunning{Radio properties of EROs}
%
\author{J. Afonso\inst{1}
\and A. M. Hopkins\inst{2}
\and M. Sullivan\inst{3}
\and B. Mobasher\inst{4}
\and A. Georgakakis\inst{5}
\and L. E. Cram\inst{6}}
\authorrunning{J. Afonso et al.}
%
%
\institute{CAAUL, Observatory of Lisbon, Tapada da Ajuda, 1349-018 Lisbon, Portugal
\and Dept. of Physics \& Astronomy, University of Pittsburgh, 3941 O'Hara St, Pittsburgh, PA 15260, USA
\and Dept. of Astronomy and Astrophysics, University of Toronto, 60 St. George St, Toronto, Ontario M5S 3H8, Canada
\and ESA, STScI, 3700 San Martin Drive, Baltimore, MD 21218, USA
\and National Athens Observatory, I.Metaxa \& Vas.Pavlou str., Athens 15236, Greece
\and Australian Research Council, GPO Box 2702, Canberra, ACT 2601, Australia}

\maketitle              

\begin{abstract}
Insensitive to dust obscuration, radio wavelengths are ideal to 
study star-forming galaxies free of dust induced biases. Using data from the 
Phoenix Deep Survey, we have identified a sample of star-forming 
extremely red objects (EROs). 
Stacking of the radio images of the radio-undetected star-forming EROs 
revealed a significant radio detection. 
Using the expected median redshift, we estimate an average 
star-formation rate of $61\,{\rm M}_{\odot}\,{\rm yr}^{-1}$ for these 
galaxies.
\end{abstract}

\section{Introduction}

Identifying and studying dusty star-forming galaxies is of vital importance
to understand galaxy evolution. Besides looking directly for the emission
from dust, these objects can also
be identified by their extreme optical to near-infrared (NIR) colours,
forming a significant fraction of the so-called Extremely Red Objects (EROs)
\cite{cimatti03}. Radio wavelengths, 
unaffected by dust obscuration and being a measure
of star formation rates in star-forming galaxies, can give an important 
insight to the characteristics of these dusty objects \cite{afonso01}. 
Performing an ERO search in an ultra-deep radio field can thus provide
an important step in understanding some of the more obscured galaxies.

\section{Star-forming EROs in the PDS}

\begin{figure}[t]
\begin{center}
\includegraphics[width=.34\textwidth]{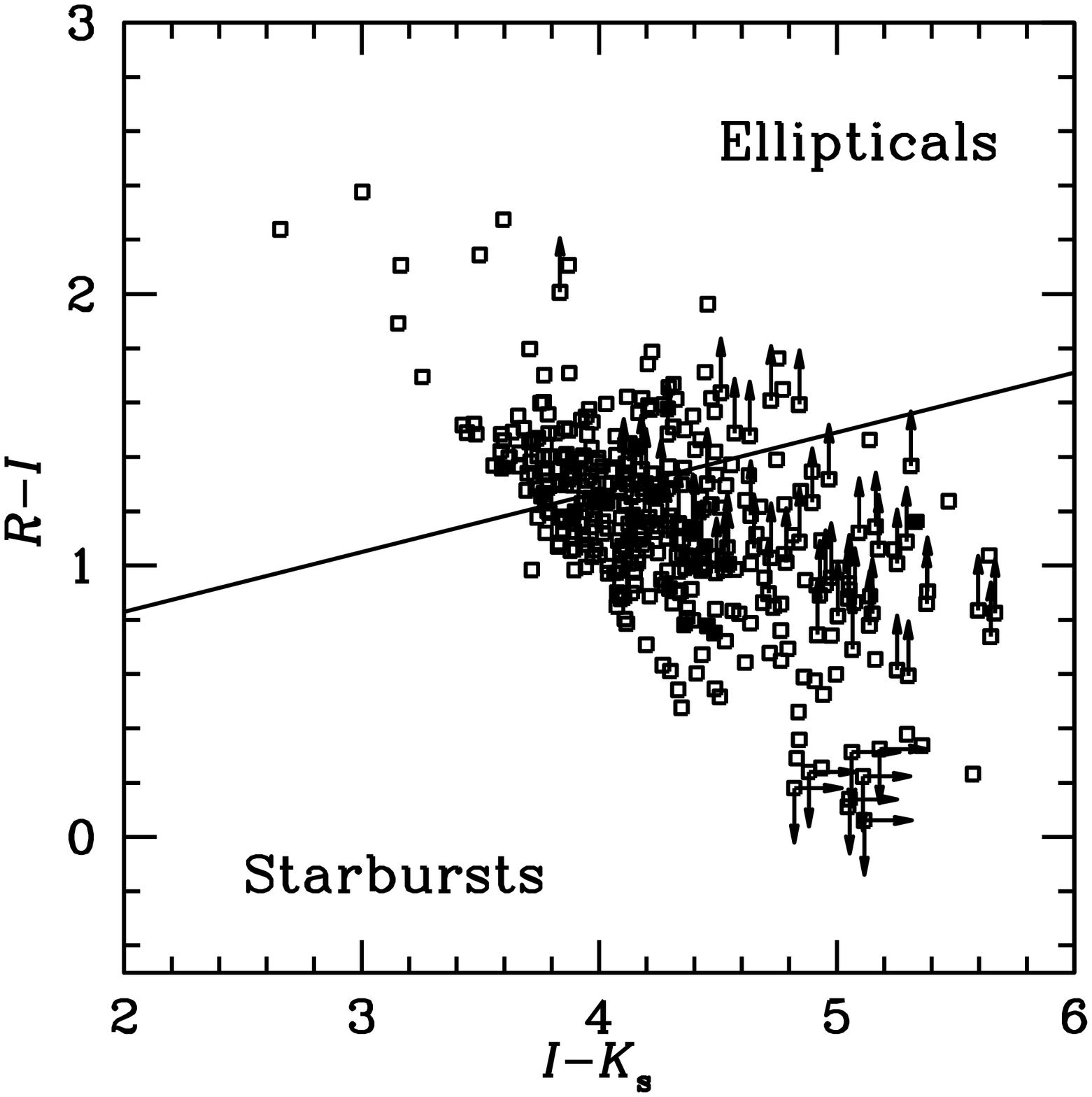}
\hspace*{0.5truecm}
\includegraphics[width=.33\textwidth]{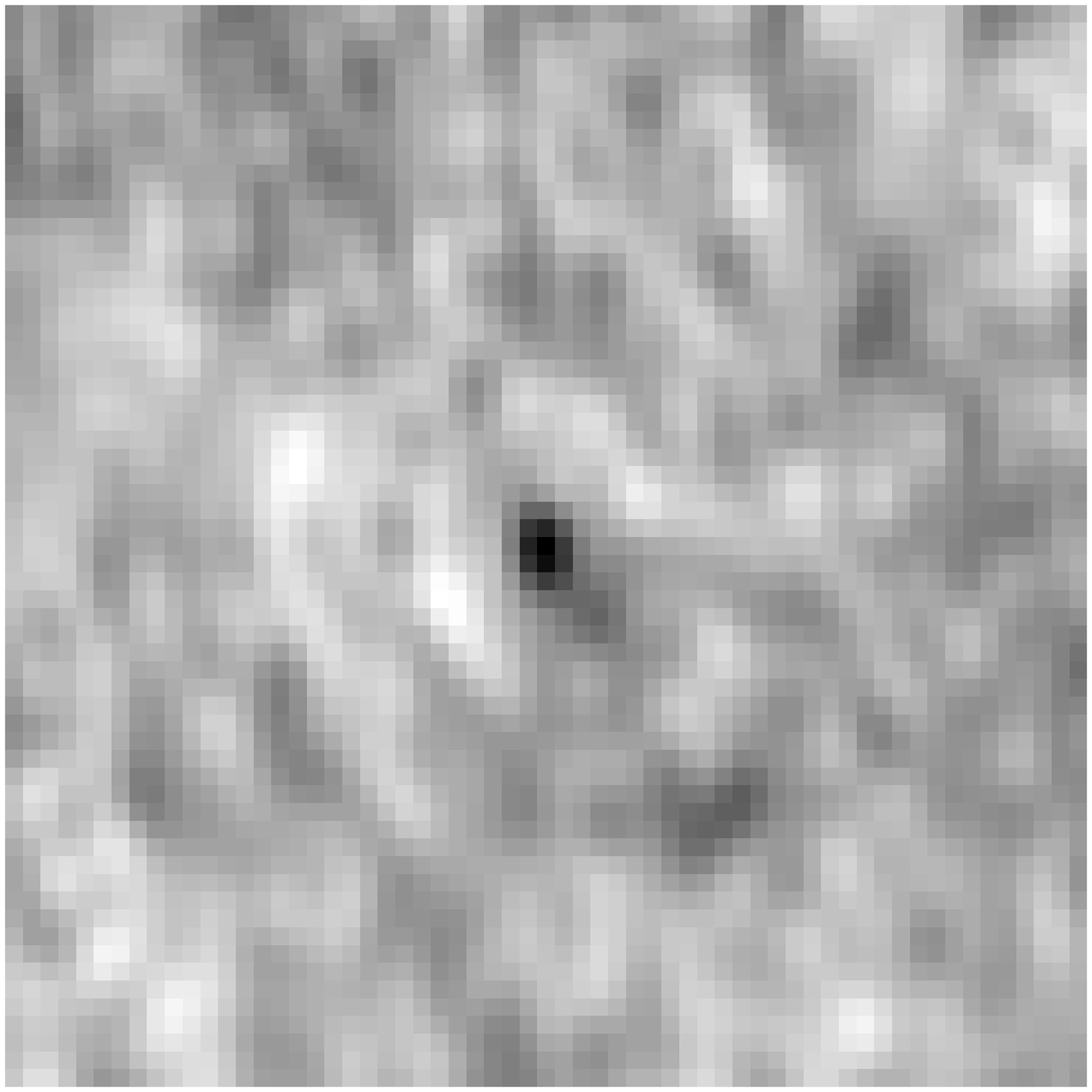}
\end{center}
\caption[]{Classification of EROs using the $RIK_{\rm s}$ colour-colour plot (left) and the stacked radio image of the radio-undetected star-forming EROs, revealing a 8.7\,$\mu$Jy detection.}
\label{plots}
\end{figure}

The Phoenix Deep Survey is a multiwavelength survey based on deep 
radio (1.4\,GHz) observations performed with the ATCA. 
With an area of over 4.5 square 
degree, and reaching flux densities of 60 $\mu$Jy, the PDS is one of the 
deepest wide angle radio surveys already made \cite{hopkins03}. Follow-up 
optical ($UBVRI$) data has recently been obtained, using the 
Wide-Field Imagers at the AAT, CTIO and ESO, reaching $\sim 25$\,mag 
in each band. Near-infrared ($K_{\rm s}$ band) observations were 
made on the central 200 square arcmin of the PDS, using SOFI at the NTT (ESO) 
to a limiting magnitude of $K_{\rm s}\sim 20$ \cite{sullivan04}. 
Based on 3$''$ aperture photometry, a sample of  
432 EROs ($R-K_{\rm s}>5$), were selected. 
Eighteen were directly identified in the radio catalogue, with radio 
fluxes between 65 and 991\,$\mu$Jy. The ERO population was separated into 
ellipticals and star-forming galaxies
using the $RIK_{\rm s}$ magnitudes, following the recent work by 
Bergstrom et al. \cite{bergstrom03}. This method seems to perform 
well for EROs with redshifts $z<1.4$ \cite{bergstrom03}, which 
seems appropriate for this survey (a spectroscopic analysis of an ERO sample
with similar $K_{\rm s}$ magnitude limits revealed a median redshift of
$z\sim 1.2$, with a very small number of EROs having $z>1.4$ \cite{cimatti03}).
A total of 214 objects fall in the star-forming region of the $RIK_{\rm s}$
colour-colour
plot (Figure~\ref{plots}). After removing all ($>3\sigma$) radio detections, 
the radio images at the locations of the star-forming 
EROs were stacked. 
A significant ($6.7\sigma$) detection of 8.7\,$\mu$Jy 
was revealed (Figure~\ref{plots}). 
Assuming a median redshift
for the sample of $z\sim 1.2$, and assuming the radio emission is solely 
due to star-formation, this implies an average star-formation rate of 
$61\,{\rm M}_{\odot}\,{\rm yr}^{-1}$. 

Improvement on the photometric analysis, by using the full photometric dataset
to estimate photometric redshifts and classifications, is currently
underway, and will result in a precise determination of the 
contribution of star-forming EROs to the star-formation rate in the Universe.

AMH acknowledges support provided by NASA through Hubble Fellowship grant HST-HF-01140.01-A awarded by STScI. JA acknowledges support from the Science and Technology Foundation (FCT, Portugal) through the fellowship BPD-5535-2001 and the research grant POCTI-FNU-43805-2001.

%

\end{document}